\documentclass[prl,aps]{revtex}
\usepackage{epsfig}
\begin{document}

\twocolumn

\bibliographystyle{prsty}

\title{Transition from Collisionless to Hydrodynamic Behavior in an Ultracold Atomic Gas}

\author{S. D. Gensemer and  D. S. Jin}

\address{JILA, Quantum Physics Division, National Institute of Standards and Technology and University of
Colorado, Boulder, Colorado 80309}

\maketitle

\begin{abstract}
Relative motion in a two-component, trapped atomic gas provides a
sensitive probe of interactions.  By studying the
 lowest frequency excitations of a two spin-state gas confined in a magnetic trap, we
  have explored the transition from the collisionless to the hydrodynamic regime.
  As a function of collision rate, we observe frequency shifts as
  large as 6\% as well as a dramatic, non-monotonic
  dependence of the damping rate.  The measurements agree qualitatively with expectations for
  behavior in the collisionless and hydrodynamic limits
 and are quantitatively compared to a classical kinetic model.
\end{abstract}

\pacs{}

%\narrowtext

Because dilute quantum gases have controllable interactions that
can be described from first principles they provide an ideal
testing ground for many-body theories of quantum fluids.
Elementary excitations, in particular, play a key role in
understanding the behavior of quantum fluids.  The nature of these
excitations varies drastically between two regimes, collisionless
and hydrodynamic, based on the relative strength of interactions.
In spite of their promise of controllable interactions, atomic gas
experiments have had difficulty exploring this full range of
behavior, in part because the high densities required to reach the
hydrodynamic regime also lead to large inelastic collision rates
and consequent rapid number loss. The recently created
Bose-Einstein condensates in metastable He, on the other hand, are
deeply in the hydrodynamic regime \cite{Robert2001,Santos2001}. In
this work we have utilized relative motion in a two-component
alkali atom gas to observe the transition from the collisionless
to the hydrodynamic regime.

In general, the underlying character of excitations, as well as
their frequencies and damping rates, depends on the relative
strength of the interactions in the system.  In the collisionless
regime, the collision rate $\Gamma_{coll}$ is much smaller than
the excitation frequency $\omega$.  In this limit, there are few
scattering events per oscillation.  Classically, the motion is
described by the single-particle Hamiltonian and collisions tend
to damp excitations.  For a quantum Fermi gas, this is the regime
of zero sound, which is a collective excitation due to the
self-consistent mean-field of a large number of particles.  The
opposite limit of large collision rate, $\Gamma_{coll}\gg\omega$,
is called the hydrodynamic or collisional regime.  Here, the
motion consists of collective excitations in which the high
collision rate maintains local equilibrium throughout the gas.
This is the regime of first sound waves, which, unlike
single-particle excitations, are only weakly damped by collisions.

For an atomic gas confined in a harmonic potential the excitation
frequency scales as the trap frequency while the collision rate
can also be controlled by varying the gas density and temperature.
Excitations of a trapped classical gas have been treated
theoretically for both the collisionless and hydrodynamic
regimes\cite{Griffin1997,Kavoulakis1998,Kavoulakis1998a,GueryOdelin1999,Bruun2000}.
However, experiments have typically been in the collisionless
regime or between the two regimes\cite{StamperKurn1998}.  In order
to reach the hydrodynamic regime at typical densities for alkali
atom experiments, we excite the lowest frequency excitation of a
two-component, magnetically confined gas.  The excitation mode
involves center-of-mass motion along the weaker, axial direction
of a cylindrically symmetric magnetic potential. Using excitations
in the weakest direction of the trap has the advantage that the
strength of the interactions relative to the excitation frequency
can be increased by tightly confining the gas radially. For a
single-component gas this dipole mode, or ``slosh," is unaffected
by interactions since collisions between atoms cannot alter
center-of-mass momentum. However, for a two-component gas,
consisting of atoms in two different spin-states for example,
collisions can impact relative motion of the two gases and this
lowest frequency mode becomes a useful probe of interactions.
Excitations of this type have been observed in studies of
two-component Bose-Einstein
condensates\cite{Maddaloni2000a,Hall1998a} and have been explored
theoretically for degenerate Fermi
gases\cite{Vichi1999,Amoruso2000}.

For this experiment the gas consisted of $^{40}$K atoms in the
internal states $f=9/2$, $m_f=9/2$ and $f=9/2$, $m_f=7/2$ (denoted
here by $\vert 9/2 \rangle$ and $\vert 7/2 \rangle$,
respectively), where $f$ is the total atomic spin and $m_f$ is the
magnetic quantum number. Because of their different magnetic
moments, atoms in these two states have slightly different
single-particle oscillation frequencies in the magnetic trap. This
fact is essential to these experiments because we find that the
crossover to the hydrodynamic regime for the slosh mode occurs at
a collision rate set by the small difference in the axial trap
frequencies for the two components.

As described in previous work\cite{DeMarco1999c}, the atoms are
precooled in a vapor-cell magneto-optical trap and then loaded
into a magnetic trap where they are further cooled by forced
evaporation. For this experiment a gas of between 0.35 and 3.5
million atoms, with a spin mixture of 45\% $\vert 9/2 \rangle$
(55\% $\vert 7/2 \rangle$), was cooled to a temperature between
0.5 and 2 $\mu$K. The magnetic trap strength corresponded to an
axial frequency of 19.84 Hz for a $\vert 9/2 \rangle$ atom. The
radial trap frequency, which was varied to access different
relative interaction strengths in the gas, was set to either 135
Hz or 256 Hz.  $^{40}$K atoms are fermions, however the
measurements described in this Letter were performed in the
classical regime by keeping the temperature of the gas above the
Fermi temperature.

To excite a slosh mode we applied an additional magnetic field
that shifted the trap center along the axial direction.  After 28
ms (roughly half of the period of oscillation in the trap) this
external field was switched off, and the motion of the atoms was
allowed to evolve freely in the magnetic trap for time $t$.  After
this delay time, the trap was switched off and the gas
ballistically expanded for 11 ms before an absorption image of the
atoms was taken. During the ballistic expansion an inhomogeneous
magnetic field was applied in order to spatially separate atoms in
the two spin states through the Stern-Gerlach effect
\cite{Stamper-Kurn1998b,DeMarco2001}. Using Gaussian fits to the
images recorded on a CCD camera we extracted the temperature,
number, and density of both components. The time evolution was
mapped out by repeating this sequence for different times $t$ and
recording the position of the center-of-mass of each component gas
after expansion.

For a spin-polarized gas of either spin state the motion of the
cloud center fit a  sine function, with no damping observed over 1
sec. The axial trap frequency was thus determined to be
$\omega_9/2\pi=19.84$ Hz for the $\vert 9/2 \rangle$ atoms and
$\omega_7/2\pi=17.44$ Hz for the $\vert 7/2 \rangle$ atoms. In
contrast, excitations of a two spin-state gas in the magnetic trap
depended strongly on collisions in the gas. Because the atoms are
fermions, s-wave collisions are forbidden between identical atoms.
Additionally, p-wave (and higher order) collisions are
energetically forbidden at the temperatures of interest
\cite{DeMarco1999c}. However, collisions can occur between atoms
in different spin states and it is these collisions that impact
relative motion in the two-component system.

\begin{figure}

\begin{center}

\epsfig{figure=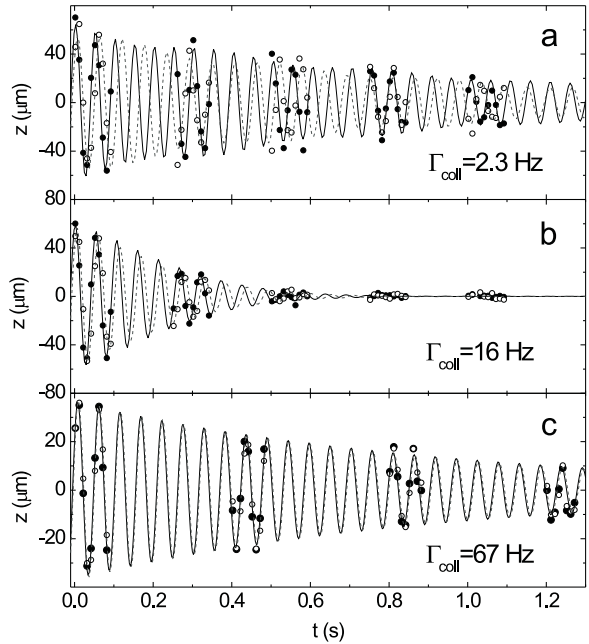,width=3.5 in, clip=}

\end{center}

\caption{Typical data showing the motion of the $\vert 9/2
\rangle$ ($\bullet$) and $\vert 7/2 \rangle$ ($\circ$) cloud
centers (z) following an abrupt shift of the trap potential.
Traces (a) and (c) correspond to the collisionless and
hydrodynamic regimes respectively, while (b) is in the transition
region. The lines are fits to a superposition of two damped
harmonic oscillator modes, with the solid and dashed lines
corresponding to the motion of the $\vert 9/2 \rangle$ ($\bullet$)
and $\vert 7/2 \rangle$ ($\circ$) cloud centers, respectively.}

\label{slosh}

\end{figure}

Typical data are shown in Fig. \ref{slosh}. On the vertical axis
are plotted the center positions of the two atom clouds, after 11
ms of free expansion.  The data in Fig. \ref{slosh}(a) is for a
small number of atoms, resulting in a relatively low collision
rate. The $\vert 7/2 \rangle$ and $\vert 9/2 \rangle$ clouds
oscillate close to their respective bare trap frequencies, but the
motion is clearly damped. In Fig. \ref{slosh}(c), we look at the
cloud motion in the hydrodynamic regime. Here a larger number of
atoms and tighter radial confinement is used to reach a much
higher collision rate. The $\vert 7/2 \rangle$ and $\vert 9/2
\rangle$ clouds now oscillate synchronously at an intermediate
frequency between the two single-particle frequencies. This
collective mode also damps, at a rate similar to the low collision
rate data in Fig. \ref{slosh}(a).  Data in Fig. \ref{slosh}(b)
show the cloud motion in the transition region between the
collisionless and hydrodynamic limits.  Here the motion of the two
clouds is coupled and the frequencies of motion are significantly
shifted from the single-particle trap frequencies.  Furthermore,
the excitation is subject to strong damping.

The lines in Fig. \ref{slosh} are fits to two modes of damped,
harmonic oscillator motion.  Because the gas has two components
the motion consists of a superposition of two normal modes, each
with an oscillation frequency $\omega$ and an exponential damping
time $\tau$.  The time-dependent center positions of the $\vert
9/2 \rangle$ and $\vert 7/2 \rangle$ gases, $z_9(t)$ and $z_7(t)$
respectively, are simultaneously fit to the following function:

\begin{center}
\begin{eqnarray}
z_9(t) &=& A_1\, e^{-t/\tau_1}\,\sin(\omega_1 t + \phi_1) + A_2\, e^{-t/\tau_2}\,\sin(\omega_2 t + \phi_2) \label{fits}\\
z_7(t) &=& B_1\, e^{-t/\tau_1}\,\sin(\omega_1 t + \theta_1) +
B_2\, e^{-t/\tau_2}\,\sin(\omega_2 t + \theta_2) \nonumber
\end{eqnarray}
\end{center}

\begin{figure}

\begin{center}

\epsfig{figure=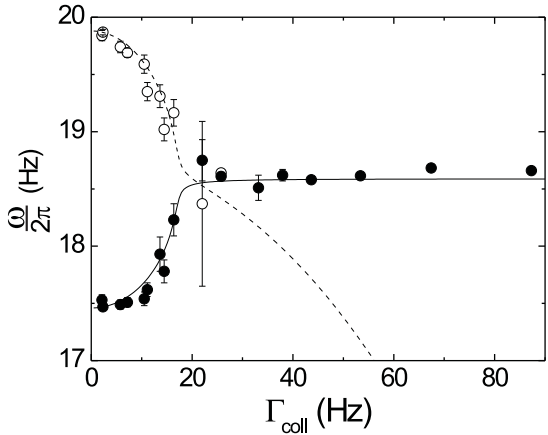,width=3.5 in, clip=}

\end{center}

\caption{Frequencies of the two excitation modes as a function of
collision rate $\Gamma_{coll}$.  The two modes frequencies
correspond to the bare trap frequencies for the two spin-states in
the limit of low collision rate. As $\Gamma_{coll}$ grows the mode
frequencies shift away from the bare trap frequencies. However a
change in behavior occurs at $\Gamma_{coll} \approx
\frac{3}{2}(\omega_9-\omega_7)$, after which only one mode, with a
frequency that is independent of $\Gamma_{coll}$, is observed.
Results of a classical kinetic model (lines) are compared to the
data.}

\label{sloshes}

\end{figure}

The measured frequencies, $\omega_1/2\pi$ and $\omega_2/2\pi$, for
different collision rates $\Gamma_{coll}$ are plotted in Fig.
\ref{sloshes}. The collision rate was varied primarily by changing
the trapped gas density, either by changing the total number of
atoms or by changing the radial trap strength. The average
collision rate per atom in the trap is $\Gamma_{coll}=2n\sigma
v/N$ where N is the total number of atoms.  The density overlap
integral $n=\int n_9(\textbf{r})n_7(\textbf{r})d^{3}\textbf{r}$
and the mean relative speed $v$ were determined from
two-dimensional Gaussian fits to the absorption images of the
expanded gas.  The collision cross-section is given by
$\sigma=4\pi a^{2}$ where the triplet scattering length for
$^{40}$K is $a=169 a_o$ ($a_o$ is the Bohr radius)\cite{Wang2000}.

As the collision rate was increased the measured frequencies
shifted from the single-particle values. This corresponds to
leaving the collisionless limit where the interactions do not
significantly affect the atomic motion. At larger $\Gamma_{coll}$,
the observed frequencies do not depend on collision rate.
Furthermore, the frequency and damping time could only be
extracted for one normal mode. This marks the transition to the
regime of hydrodynamic or first sound where $\Gamma_{coll}$ is
high enough to give rise to a collective excitation in the
classical gas. As can be seen in Fig. \ref{slosh}(c) the observed
collective mode corresponds to in-phase motion of nearly equal
amplitude for the two component gases. The frequency of this mode
lies between the two bare frequencies with a shift of
approximately $6\%$ compared to either $\omega_9$ or $\omega_7$.

\begin{figure}

\begin{center}

\epsfig{figure=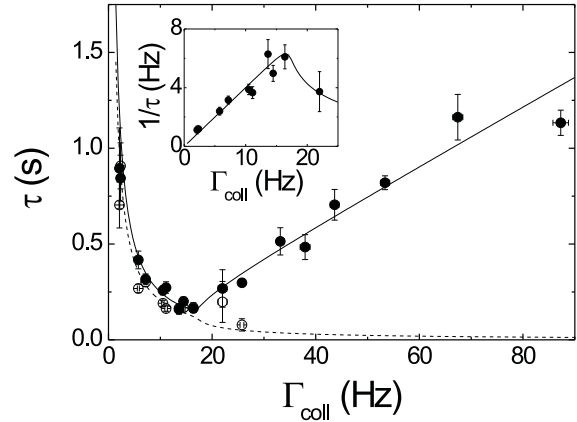,width=3.5 in, clip=}

\end{center}

\caption{Exponential damping times for the two modes of
excitation. The damping times exhibit a minimum at the transition
from the collisionless to the hydrodynamic regime. In the
hydrodynamic regime, the measured $\tau$ is seen to depend
linearly on $\Gamma_{coll}$.  The inset shows the damping rate
$1/\tau$ which scales linearly with $\Gamma_{coll}$ in the
collisionless regime.}

\label{damping}

\end{figure}

The measured exponential damping times, $\tau_1$ and $\tau_2$, are
plotted in Fig. \ref{damping}.  We observe a minimum in $\tau$ (or
correspondingly a maximum in the damping rate) near the point
where collective mode emerges. This maximum marks the transition
between the collisionless and hydrodynamic regimes and is
well-known in the context of quantum fluids\cite{Pines1966}. In
this transition regime the collisions are strong enough to impede
single-particle excitations but are not sufficient to give rise to
a well-defined sound wave. As expected the damping rate $1/\tau$
increases linearly with $\Gamma$ in the collisionless regime (see
inset to Fig. \ref{damping}) and decreases as $1/\Gamma$ in the
hydrodynamic regime.

In Figs. \ref{sloshes} and \ref{damping} we also compare the data
to the results of a classical kinetic model.  This model
reproduces the previous results by Vichi and Stringari
\cite{Vichi1999} but can also accommodate different trap
frequencies for the two spin-states as well as different numbers
of atoms in the two components.  We model the system with the
following coupled equations of motion:

\begin{center}
\begin{eqnarray}
        {\ddot{z_9}} &=& \frac{-F_d}{N_9~m} - {\omega_9}^2~z_9   \\
        {\ddot{z_7}}&=& \frac{F_d}{N_7~m} - {\omega_7}^2~z_7  , \nonumber
\end{eqnarray}
${\rm where}\; F_d =\frac{1}{3}
m~N~\Gamma_{coll}~(\dot{z_9}-\dot{z_7}).\nonumber$

\end{center}

Here $N_9$ and $N_7$ are the numbers of atoms in each spin-state,
and $m$ is the atom mass.  The viscous damping force due to the
collisional interactions in the gas, $F_d$, was be derived by
considering the effect of individual collision events on the
cloud's center-of-mass motion and integrating over all possible
collisions. The expression $F_d$ assumes a classical gas and a
small amplitude for the slosh.  The solution of these equations
consists of two normal modes whose frequencies and exponential
damping times are shown as the lines in Figs. \ref{sloshes} and
\ref{damping}. For these theory lines the bare trap frequencies,
as well as the measured temperature, number, and spin composition
(45\% $\vert 9/2 \rangle$), are input to the model. In addition,
we allow a single free parameter that is a multiplicative factor
that scales the collision rate axis. The best fit scaling
corresponds to 1.30$\pm$0.02, which implies that the
experimentally determined collision rates are low.  This scaling
factor is consistent with our estimated $\pm$50\% systematic
uncertainty in extracting $N$ from absorption images.

The model gives excellent agreement with the data for both
frequency shifts and damping times.  The model reveals the
existence of a second, strongly damped collective mode in the
hydrodynamic regime.  This collective mode corresponds to the
spin-dipole mode studied by Vichi and Stringari for a degenerate
Fermi gas \cite{Vichi1999} and is overdamped in the hydrodynamic
regime. This type of excitation is also observed as the giant
dipole resonance in nuclei. The model can also answer the question
of what sets the scale for the transition from the collisionless
to the hydrodynamic regime in the two-component system.  Varying
the bare trap frequencies in the model reveals that the maximum in
the damping rate, which marks the emergence of hydrodynamic
behavior, scales with the frequency difference of the bare
modes\cite{note}.

\begin{figure}

\begin{center}

\epsfig{figure=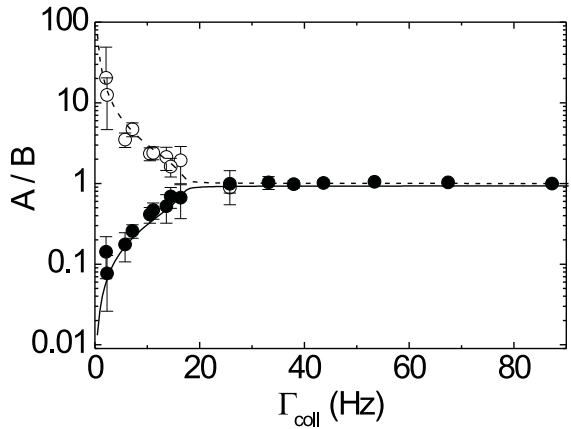,width=3.5 in, clip=}

\end{center}

\caption{Ratio of the amplitudes for $\vert 9/2 \rangle$ and
$\vert 7/2 \rangle$ motion.  The ratios of the amplitudes,
$A_1/B_1$ and $A_2/B_2$, are plotted versus collision rate
$\Gamma_{coll}$.  As $\Gamma_{coll}$ increases the motion of the
two gases becomes increasingly coupled until a collective mode,
consisting of equal amplitude motion of the two gases, emerges in
the hydrodynamic regime.}

\label{amplitudes}

\end{figure}

Finally in Fig. \ref{amplitudes} we show the emergence of coupled
motion in the two-component gas by plotting the amplitude ratios
$A_1/B_1$ and $A_2/B_2$ extracted from the fits to Eqn. 1. Each
normal mode is in general a linear combination of motion of the
two component gases.  A finite ratio corresponds to a nontrivial
combination and Fig. \ref{amplitudes} reveals that the motion of
the two species becomes coupled as $\Gamma_{coll}$ increases.

In conclusion, by exploiting relative motion in a two-component
trapped atom gas we have examined the transition from the
collisionless to the hydrodynamic regime.   The  ability to access
the full range of excitation behaviors demonstrated here will be
extremely useful for studying elementary excitations of quantum
gases, whether Bose-Einstein condensates or Fermi degenerate
gases.  In the Fermi system, for example, quantum statistical
reduction in collision rate could be revealed in a study of
excitations in the hydrodynamic regime\cite{Bruun1999}. Although
it has been previously suggested that this regime would be
difficult to reach for $^{40}$K
experiments\cite{Vichi1999,Bruun1999}, the results of this work
prove otherwise. Excitations in the hydrodynamic regime could also
be used to reveal the onset of superfluidity
\cite{Bruun2000a,Baranov2000}. In addition, spin excitations such
as described here are interesting in their own right for quantum
degenerate gases.

We thank B. DeMarco, E.A. Cornell and J.L. Bohn for helpful
discussions.  This work is supported by the National Science
Foundation, the Office of Naval Research, and the National
Institute of Standards and Technology.

\bibliography{hydrodynamic}

\begin{thebibliography}{10}
\bibitem{note}
In the limit of degenerate bare trap frequencies, the in-phase collective mode
  is no longer subject to collisional damping.
\end{thebibliography}

\end{document}